\documentstyle[preprint,revtex,eqsecnum]{aps}
\begin{document}
\draft

\hyphenation{
mini-su-per-space
pre-factor
mani-fold
mani-folds
}


\begin{title}
No-boundary theta-sectors in \\
spatially flat quantum cosmology
\end{title}
\author{Domenico Giulini\cite{nico}}
\begin{instit}
Institut f\"ur Theoretische Physik,
Universit\"at Freiburg,\\
Hermann-Herder Strasse 3, D-W-7800 Freiburg, Germany
\end{instit}
\author{Jorma Louko\cite{jorma}}
\begin{instit}
Department of Physics,
Syracuse University,\\
Syracuse, New York 13244--1130, USA
\end{instit}

\begin{abstract}
Gravitational theta-sectors are investigated in spatially locally
homogeneous cosmological models with flat closed spatial surfaces
in 2+1 and 3+1 spacetime dimensions. The metric ansatz is kept in
its most general form compatible with Hamiltonian minisuperspace
dynamics. Nontrivial theta-sectors admitting a semiclassical
no-boundary wave function are shown to exist only in 3+1
dimensions, and there only for two spatial topologies. In both
cases the spatial surface is nonorientable and the nontrivial
no-boundary theta-sector unique. In 2+1 dimensions the
nonexistence of nontrivial no-boundary theta-sectors is shown to
be of  topological origin and thus to transcend both the
semiclassical approximation and the minisuperspace ansatz.
Relation to the necessary condition given by Hartle and Witt for
the existence of no-boundary theta-states is discussed.
\end{abstract}
\pacs{Freiburg THEP-92/7 \\ Syracuse SU-GP-92/2-3}

\narrowtext

\section{Introduction}
\label{sec:introduction}

It is well known that the quantization of classical theories
with topologically nontrivial configuration spaces may display
more variety than what is present in theories built from
topologically trivial configuration
spaces~\cite{isham/houches,jackiw,fried/sorkin,ishamPL}. This is
due to the fact that the ambiguities present in most quantization
prescriptions emerge as additional freedom once the classical
configuration space meets certain specific topological conditions.
A convenient heuristic approach for classifying this additional
freedom is the functional Schr\"odinger representation, in which
the wave function is taken to be a section in a possibly
nontrivial complex line bundle over the classical configuration
space~\cite{isham/houches,jackiw}. The ambiguities are then
precisely measured by the well known classification of flat
complex line bundles with connection, which is simply given by the
inequivalent, irreducible one-dimensional representations of the
fundamental group of the classical configuration space. A familiar
example is provided by the theta-sectors in Yang-Mills
theories~\cite{jackiw}, where the fundamental group of the
underlying classical configuration space is~{\bf Z}. The
representations there are thus labeled just by an $S^1$-valued
quantity (forming the character group), called the theta-angle. It
has become customary to extend the term ``theta-sector" to all
cases where a group acts as redundancy-symmetries on a
configuration space; ``theta" then just refers to the quantity
that labels the inequivalent, irreducible, one-dimensional
representations of this group carried by the wave function.
Clearly, it may be continuous or discrete.

One theory in which theta-sectors can occur is canonically
quantized gravity, both in the spatially open and closed
case~\cite{ishamPL}. In the latter case the classical
configuration space is
$\hbox{Riem($\Sigma$)}/\hbox{Diff($\Sigma$)}$, or the space of
Riemannian geometries on a (closed) three-manifold~$\Sigma$. The
momentum constraint equations \cite{dewitt} imply that the wave
functions are invariant under~\hbox{Diff$_0$($\Sigma$)}, the
connected component of~\hbox{Diff($\Sigma$)}. If
\hbox{Diff($\Sigma$)} is not connected, there is a freedom in how
the wave function is to transform under the disconnected
components of~\hbox{Diff($\Sigma$)}. How to secure that the
physical predictions of the theory are coordinate-independent in
the presence of disconnected diffeomorphisms may depend on how one
anticipates ``physical predictions" to arise from canonically
quantized gravity; however, experience with ordinary quantum
mechanics and quantum field theory suggests that a sufficient
condition would be that the wave function transform according to
a one-dimensional, irreducible representation of
$\hbox{Diff($\Sigma$)}/\hbox{Diff$_0$($\Sigma$)}
=\hbox{$\pi_0\bigl({\rm Diff}(\Sigma)\bigr)$}$. If
\hbox{Diff($\Sigma$)} acted freely on \hbox{Riem($\Sigma$)},
\hbox{$\pi_0\bigl({\rm Diff}(\Sigma)\bigr)$} would be homeomorphic
to the fundamental group of
$\hbox{Riem($\Sigma$)}/\hbox{Diff($\Sigma$)}$ and a discussion
could just be given along the general lines indicated above. For
closed cosmologies, however, the action of the relevant
diffeomorphism group  on \hbox{Riem($\Sigma$)} is not free. It
fixes the metrics with isometries. The analogy to the standard
arguments using the fundamental group breaks down at this point.
But, as already emphasized, it is still useful to talk of theta
sectors associated to one-dimensional, irreducible
representations of a group acting on the configuration space, even
if the action is not free. For a simpler model such a situation
has been comprehensively studied in Ref.~\cite{hajicek/bi} using
canonical quantization.

Wave functions satisfying the constraint equations of canonically
quantized gravity can be formally obtained from path integral
constructions that satisfy sufficient conditions of local
invariance~\cite{hallhartle}. To recover from path integrals wave
functions that would also have the desired transformation
properties under the disconnected components
of~\hbox{Diff($\Sigma$)}, one may need to generalize the integral
to explicitly include an appropriate sum. In the context of
non-relativistic quantum mechanics a generalization of this kind
has first been described in Ref.~\cite{laid/dewitt}. In quantum
cosmology and for the particular case of the no-boundary
prescription of Hartle and Hawking \cite{hh,hartle/car} an
analogous discussion has been given by Hartle and
Witt~\cite{hartle/witt}, who showed that an ambiguity under the
disconnected components of \hbox{Diff($\Sigma$)} may emerge in
this prescription as well. It is the purpose of the present paper
to investigate this ambiguity in the no-boundary prescription
within a family of simple cosmological models.

To be more specific, recall that the basic building block of the
no-boundary wave function is~\cite{hh,hartle/car,hartle/witt}
\begin{equation}
{\Phi_{N\!B}} ({h_{ij}}; \Sigma)
=
\int {\cal D} {g_{\mu\nu}} \exp \bigl[ - I({g_{\mu\nu}};{\cal M})
\bigr]
\label{phihh}
\end{equation}
where ${\cal M}$ is a compact four-dimensional manifold with
boundary~$\Sigma$, and the functional integral is over some
appropriate (presumably complex~\cite{hallhartle/contour}) class
of metrics ${g_{\mu\nu}}$ on ${\cal M}$ such that the induced
three-metric on $\Sigma$ is ${h_{ij}}$ as specified in the argument
of~${\Phi_{N\!B}}$. More generally, one could consider a sum of
terms of this form over several four-manifolds with some
appropriate weights. Let us assume that the measure
${\cal D}{g_{\mu\nu}}$ can be and has been chosen to be invariant
under all diffeomorphisms of~${\cal M}$, both connected and
disconnected. This will guarantee the invariance of
${\Phi_{N\!B}}$ under those diffeomorphisms of $\Sigma$ that are
restrictions of a diffeomorphism of~${\cal M}$. However, if there
exist diffeomorphisms of $\Sigma$ that are not restrictions of any
diffeomorphism of~${\cal M}$, ${\Phi_{N\!B}}$ need not have any
simple transformation property under such diffeomorphisms. In such
cases it is always possible to construct a wave function that is
strictly invariant under \hbox{Diff($\Sigma$)} by formally summing
${\Phi_{N\!B}}$ over~\hbox{$\pi_0\bigl({\rm
Diff}(\Sigma)\bigr)$}. However, it may also be possible to form
from ${\Phi_{N\!B}}$ sums that transform under nontrivial
representations of~\hbox{$\pi_0\bigl({\rm Diff}(\Sigma)\bigr)$}.
Hence there exists the possibility of no-boundary theta-states.

In this paper we shall address this question within a class of
spatially homogeneous minisuperspace models~\cite{ryan/shepley}.
Although the relevance of minisuperspace quantization for the full
quantum theory may be debatable~\cite{kuchar/ryan}, such models
have been extensively invoked as a qualitative arena for quantum
gravity. What interests us here is that there exist
minisuperspace models that both possess theta-sectors in the
canonical quantization and also admit wave functions compatible
with the no-boundary prescription. It is perhaps appropriate to
first illustrate this in a simple example.

Consider a model defined in three spacetime dimensions by the
(Lorentzian) metric ansatz
\begin{equation}
ds^2 =
- N^2 (t) dt^2
+
a^2 (t) dx^2
+
b^2 (t) dy^2
\label{exansatz}
\end{equation}
where the coordinates $x$ and $y$ are periodic with
period~$2\pi$. The wave functions in canonical quantization
depend on the two scale factors $a$ and~$b$, and they obey the
minisuperspace Wheeler-DeWitt equation~\cite{louko/tuckey}
\begin{equation}
\left(
{\partial^2 \over \partial a \, \partial b}
+
{\pi^2 \Lambda \over 4 G^2} \,
ab \right)
\Psi = 0
\label{exwdw}
\end{equation}
where $G$ and $\Lambda$ are respectively the gravitational
constant and the cosmological constant (we use units in which
$\hbar=c=1$). Generic solutions to (\ref{exwdw}) need have no
particular symmetry under interchanging $a$ and~$b$. However,
interchanging $a$ and $b$ corresponds to a disconnected
diffeomorphism which permutes the two $S^1$ factors in the metric
ansatz~(\ref{exansatz}). The minisuperspace analogue of
\hbox{$\pi_0\bigl({\rm Diff}(\Sigma)\bigr)$} in this model is
therefore the permutation group of the two scale factors, which
has exactly two one-dimensional irreducible representations. The
trivial theta-sector consists of wave functions that are symmetric
in $a$ and~$b$, and the nontrivial sector of wave functions that
are antisymmetric in $a$ and~$b$.

We can now ask whether the two theta-sectors in this model contain
wave functions that could arise from the no-boundary proposal.
Here, and in most of this paper, we shall only address this
question at the semiclassical level (for discussions on defining a
minisuperspace no-boundary type wave function beyond the
semiclassical level, see
Refs.~\cite{wy,loukoPL,hall/louko3,hayw/louko,garay,louko/whit}):
we ask whether the two theta-sectors contain wave functions that
have the semiclassical form expected of a no-boundary wave
function. For this, recall that the classical Euclidean actions
with the no-boundary data in this model are given by the
formula~\cite{louko/tuckey,spatflat1}
\begin{equation} I^c =
\mp
{\pi b \sqrt{1 - \Lambda a^2}
\over 2G}
\ \ ,
\label{exaction}
\end{equation}
which by itself is not symmetric in $a$ and~$b$, and by a similar
formula with $a$ and $b$ interchanged. Let now $\Psi_0(a,b)$ be a
solution to (\ref{exwdw}) such that its semiclassical form
involves one or both of the actions~(\ref{exaction}). (Such
solutions do exist~\cite{louko/tuckey}.)  The wave functions
defined by
$
\Psi_\pm (a,b) =
\Psi_0(a,b) \pm
\Psi_0(b,a)
$
are then respectively symmetric and antisymmetric under
permutations of $a$ and~$b$, and their semiclassical form involves
the classical actions of the no-boundary solutions. Hence
$\Psi_\pm$ can be understood as no-boundary wave functions in
respectively the trivial and nontrivial theta-sectors.

The models we shall consider are the locally spatially homogeneous
cosmological models with flat spatial sections, both in 2+1 and
3+1 spacetime dimensions, such that the metric ansatz is kept in
its most general form compatible with Hamiltonian minisuperspace
dynamics. (Note that the model exhibited above does not belong to
this class. The corresponding general model with $S^1\times S^1$
spatial topology allows a nondiagonal metric and will be discussed
in Section~\ref{sec:2+1}.) The classical (Euclidean) solutions
with the no-boundary boundary data in these models were found in
Ref.~\cite{spatflat1}, and also some discussion of the
minisuperspace analogues of \hbox{$\pi_0\bigl({\rm
Diff}(\Sigma)\bigr)$} was given. Although there are five models in
which the minisuperspace analogue of \hbox{$\pi_0\bigl({\rm
Diff}(\Sigma)\bigr)$} acts nontrivially on the semiclassical
no-boundary contributions to the wave function, we shall show that
nontrivial no-boundary theta-sectors exist only in 3+1 dimensions,
and there only for two spatial topologies. In both cases the
spatial surface is non-orientable, and the nontrivial no-boundary
theta-sector is unique.

In 2+1 dimensions, we shall show that the absence of nontrivial
no-boundary theta-sectors has a topological origin, independent
from our having arrived at this result within minisuperspace. In
this situation the three-dimensional version of ${\Phi_{N\!B}}$
(\ref{phihh}) is not by itself invariant under all of
\hbox{$\pi_0\bigl({\rm Diff}(\Sigma)\bigr)$}, owing to the
existence of non-extendible diffeomorphisms of~$\Sigma$; however,
the extendible and non-extendible diffeomorphisms are intertwined
in a way which renders a trivially transforming sum of
${\Phi_{N\!B}}$'s the only possibility. This example thus shows
that the necessary condition found by Hartle and Witt for the
existence of nontrivial no-boundary theta-sectors is not a
sufficient one.

The paper is organised as follows. In
Section~\ref{sec:minidiffeos} we introduce the models, establish
the notation, and define the minisuperspace analogue
of~\hbox{$\pi_0\bigl({\rm Diff}(\Sigma)\bigr)$}. The 2+1- and
3+1-dimensional models are analyzed respectively in Sections
\ref{sec:2+1} and~\ref{sec:3+1}. Section~\ref{sec:summary}
contains a summary and discussion. Some technical points are
postponed to the appendices.

\section{Minisuperspace spatial diffeomorphisms
and the no-boundary wave function}
\label{sec:minidiffeos}

We consider 2+1- and 3+1-dimensional cosmological models whose
Lorentzian metric ansatz is given by
\begin{equation}
ds^2 = - N^2(t) dt^2
+ {h_{ij}} (t)  dx^i dx^j
\label{ansatz}
\end{equation}
where $\{x^i\}$ are a set of two (three, respectively) local
spatial coordinates. The spatial surfaces are taken to be closed
(compact without boundary). The flatness of the spatial metric
then implies that only finitely many spatial topologies are
possible: two in 2+1 dimensions and 10 in 3+1
dimensions~\cite{wolf}. Both in 2+1 and 3+1 dimensions one of
these topologies is the torus  $S^1\times S^1$ ($S^1\times
S^1\times S^1$), and the remaining ones can be obtained from the
torus by identifications under a discrete group. We choose the
coordinates $\{x^i\}$ to be a set of angle coordinates, each
periodically identified by~$2\pi$. For the toroidal spatial
topology these periodic identifications are the only ones, and
${h_{ij}}$ is a symmetric positive definite matrix with no other
constraints on its three (six) independent components. For the
non-toroidal spatial topologies there are further identifications
in the coordinates, as well as linear relations between the
components of ${h_{ij}}$~\cite{spatflat1,wolf}. A Hamiltonian
formulation of the dynamics is obtained by performing the spatial
integration in the Einstein action~\cite{einstein,york,jantzen}.

The minisuperspace analogue of a spatial diffeomorphism is now a
transformation
\begin{equation}
x^i \>=\> {M^i{}_j} \,
{\bar x}^j
\label{minidiffeo}
\end{equation}
where ${M^i{}_j}$ is a $2\times2$ ($3\times3$) matrix such that
the global identifications look identical when expressed in the
``old" coordinates $\{x^i\}$ and in the ``new"
coordinates~$\{{\bar x}^i\}$.  We shall call the group formed by
such matrices the minisuperspace spatial diffeomorphism group and
denote it by~$G$. Since for all the spatial topologies the
identifications include periodic identifications by~$2\pi$, $G$ is
a subgroup of \hbox{${\it GL}(2,{\bf Z})$} (\hbox{${\it GL}(3,{\bf
Z})$}). In the present context and for orientable spatial surfaces
one may or may not in addition require that the diffeomorphisms
preserve the orientation, which implies $\det(M)=1$ thus leading
to a subgroup of \hbox{${\it SL}(2,{\bf Z})$} (\hbox{${\it
SL}(3,{\bf Z})$}). It turns out, however, that the results we are
after are independent of such a requirement.

The group $G$ acts on the metric components by
\begin{equation}
{\bar h}_{ij}
\;=\;
{(M^T h M)}_{ij}
\ \ ,
\label{minidiffeoh}
\end{equation}
where we have adopted a matrix notation for the summation over
the indices. We now require that the wave function transform
according to
\begin{equation}
\Psi \left(
{(M^T h M)}_{ij}
\right)
\;=\;
\chi(M) \,
\Psi \left( {h_{ij}} \right)
\ \ ,
\label{psitransform}
\end{equation}
where $\chi$ is a one-dimensional representation of the group
$\tilde G$ formed by the transformations~(\ref{minidiffeoh}). This
is the minisuperspace analogue of the condition that the wave
function of the full theory transform according to a
one-dimensional representation of \hbox{$\pi_0\bigl({\rm
Diff}(\Sigma)\bigr)$}. Note that, in general, $\tilde G$ may only
be a factor group of~$G$.

For those spatial topologies that admit classical solutions with
the no-boundary data, the no-boundary wave function is expected to
take the semiclassical form
\begin{equation}
\Psi_{N\!B} \left({h_{ij}}\right)
\sim
\sum_k P_k\left({h_{ij}}\right)
\exp \bigl[- I_k\left({h_{ij}}\right) \bigr]
\label{semiclnb}
\end{equation}
where $I_k\left({h_{ij}}\right)$ are the actions of the classical
solutions as functions of the boundary data~${h_{ij}}$,
$P_k\left({h_{ij}}\right)$ are the semiclassical prefactors, and
$k$ is the index labelling the classical solutions. As the set
$\bigl\{I_k\left({h_{ij}}\right)\bigr\}$ must by construction be
closed under the action of $G$~(\ref{minidiffeoh}), it is always
possible to choose the prefactors so that (\ref{semiclnb})
transforms under the trivial representation of~$\tilde G$. There
may even be several such choices for the prefactors, if the action
of $G$ in the set  $\bigl\{I_k\left({h_{ij}}\right)\bigr\}$ is not
transitive. Our purpose is to determine, using the explicitly
known forms of $I_k\left({h_{ij}}\right)$~\cite{spatflat1},
whether the prefactors can be assigned so that (\ref{semiclnb})
transforms under a nontrivial representation of~$\tilde G$.

We shall make two assumptions. Firstly, we assume that for each
$k$ the prefactor $P_k\left({h_{ij}}\right)$ is slowly varying
compared with $\exp \bigl[- I_k\left({h_{ij}}\right) \bigr]$ in
the usual semiclassical sense. Secondly, we assume that the terms
in (\ref{semiclnb}) coming from the different $k$'s can be regarded
as a linearly independent set. In the models where the sum is
finite, the first assumption implies the second one. In the more
interesting models where there are a countable infinity of
solutions to the classical boundary value problem the second
assumption is nontrivial; however, even in these models the first
assumption can be shown to imply linear independence of every
finite subset of the terms. In the simplest model with a countable
infinity of semiclassical contributions we shall demonstrate this
in Appendix~\ref{app:linear}.

For simplicity, we take the cosmological constant to be vanishing,
and we take the Einstein action~\cite{einstein,york} to be
defined with the conventional positive sign of $\sqrt{g}$ for
positive definite metrics. (Motivations for considering the
unconventional sign of $\sqrt{g}$ in the no-boundary proposal are
discussed in Ref.~\cite{hallhartle/contour}.)  Allowing a
nonvanishing cosmological constant and a sign ambiguity in
$\sqrt{g}$ would however not bring anything new to the
transformation laws of the classical actions under $\tilde
G$~\cite{spatflat1}, and the conclusions about the transformation
properties of semiclassical no-boundary wave functions would
remain the same.

\section{2+1 dimensions: $S^1\times S^1$ spatial surfaces}
\label{sec:2+1}

In 2+1 dimensions the only possible spatial topologies are the
two-torus and the Klein bottle~\cite{wolf}. With both topologies
there exist classical solutions with the no-boundary data, but in
the Klein bottle case the minisuperspace spatial diffeomorphisms
act trivially on the classical actions~\cite{spatflat1}. We
therefore only need to consider the two-torus case.

For the two-torus, the only identifications in the coordinates
$\{x^i\}=\{x,y\}$ are the periodic ones,
\begin{equation}
(x,y) \sim (x+2\pi,y) \sim (x,y+2\pi)
\ \ ,
\label{twotorusids}
\end{equation}
and ${h_{ij}}$ is a symmetric positive definite $2\times2$ matrix
with three independent components. As the two-torus is orientable,
we choose to restrict attention to orientation-preserving
minisuperspace spatial diffeomorphisms which, as we will soon see,
implies no loss of generality. We thus have $G=\hbox{${\it
SL}(2,{\bf Z})$}$. In the action of $G$ on ${h_{ij}}$
by~(\ref{minidiffeoh}), the only degeneracy is in the overall sign
of~$M$. Hence ${\tilde G}=\hbox{${\it PSL}(2,{\bf
Z})$}\equiv\hbox{${\it SL}(2,{\bf Z})$}/\{1,-1\}$.

Let us first list all one-dimensional representations
of~\hbox{${\it PSL}(2,{\bf Z})$}. Recall~\cite{coxeter,rankin}
that \hbox{${\it SL}(2,{\bf Z})$} is generated by the two matrices
\begin{equation}
A = \pmatrix{
1&1\cr
0&1\cr}
\quad,\quad
B=\pmatrix{1&0\cr1&1\cr}\quad
\label{sltwogenerators1}
\end{equation}
or, equivalently, by the two matrices
\begin{equation}
\begin{array}{rcl}
S &:=& B^{-1}A=\pmatrix{1&1\cr-1&0\cr} \\
\noalign{\vskip2\jot}
T &:=& B^{-1}AB^{-1}=\pmatrix{0&1\cr-1&0\cr}\,,
\end{array}
\label{sltwogenerators2}
\end{equation}
whose only independent relations are
\begin{equation}
S^3 = T^2 = -1\,.
\label{sltworelations}
\end{equation}
The group \hbox{${\it PSL}(2,{\bf Z})$} can be generated by the
same matrices if each matrix is understood modulo the overall
sign. In particular, the minus sign in the relations
(\ref{sltworelations}) then disappears so that the group
\hbox{${\it PSL}(2,{\bf Z})$} may be identified with the free
product, ${\bf Z}_3*{\bf Z}_2$, of ${\bf Z}_3$ (generated by~$S$)
and ${\bf Z}_2$ (generated by~$T$). Its abelianization is
therefore ${\bf Z}_3\times {\bf Z}_2\cong {\bf Z}_6$ so that we
have a total of six one-dimensional representations $\chi$,
uniquely characterized by the independent choices of
$\chi(S)\in\{1, e^{2\pi i/3},  e^{-2\pi i/3} \}$ and
$\chi(T)\in\{1, -1 \}$.

The actions of the classical solutions with the no-boundary data
are labelled by ordered pairs of coprime integers, modulo the
overall sign of the pair. We shall denote the equivalence classes
of such pairs modulo the overall sign by~${(m,n)}$. The explicit
formula is~\cite{spatflat1}
\FL
\begin{equation}
I_{(m,n)} \left({h_{ij}}\right)
\>=\>
- {\pi \over 2G}
\,
\sqrt{h\over
{m^2h_{xx}+
n^2h_{yy}+
2mnh_{xy}}
}
\label{Imn}
\end{equation}
where
$h=\det\left({h_{ij}}\right)=
h_{xx}h_{yy}-{\left(h_{xy}\right)}^2$. These are the actions to be
used in~(\ref{semiclnb}), and the summation label $k$ is thus
replaced by ${(m,n)}$. As mentioned in
Section~\ref{sec:minidiffeos}, we assume that the semiclassical
contributions in (\ref{semiclnb}) can be treated as a countable
linearly independent set. In Appendix \ref{app:linear} it will be
shown that every finite subset of these contributions is linearly
independent by virtue of the assumption that the prefactors are
slowly varying compared with the exponential factors.

On the set of the classical actions~(\ref{Imn}), the \hbox{${\it
PSL}(2,{\bf Z})$} transformations act by
\begin{equation}
I_{(m,n)}
\left(
{(M^T h M)}_{ij}
\right)
=I_{{(m,n)}\cdot M^{T}}\left({h_{ij}}\right)\,.
\label{Itransform}
\end{equation}
This \hbox{${\it PSL}(2,{\bf Z})$} action is obviously
transitive. By the assumption of linear independence, the
transformation law~(\ref{psitransform}) then implies that the
prefactors must split into the form
\begin{equation}
P_{(m,n)}\left({h_{ij}}\right) = \sigma_{(m,n)}\
\pi_{(m,n)}\left({h_{ij}} \right)
\label{prefactorsplit}
\end{equation}
where $\pi_{(m,n)}\left({h_{ij}}\right)$ transforms in the same
way as $I_{(m,n)}\left({h_{ij}}\right)$ in equation
(\ref{Itransform}) and the numerical factors $\sigma_{(m,n)}$
transform according to
\begin{equation}
\sigma_{{(m,n)}\cdot M^{T}}=\chi (M^{-1})\sigma_{(m,n)}\,.
\label{sigmatransform}
\end{equation}
Note that as a consequence of the covariance of the Wheeler-DeWitt
equation the transformation property for $\pi_{(m,n)}$ is
consistent with the requirement that the total wave function is a
solution to the Wheeler-DeWitt equation.

Let us now discuss which of the six representations $\chi$ are
compatible with~(\ref{sigmatransform}). For this observe that if
$M$ possesses a fixed point, $(m_0,n_0)$, in the sense
$(m_0,n_0)\cdot M^{T}=(m_0,n_0)$, then $\chi$ must represent $M$
trivially. Suppose $H\subset\hbox{${\it PSL}(2,{\bf Z})$}$ is a
set of $M$'s, each of which possesses at least one fixed point,
and let $N_H$ be the smallest normal subgroup of \hbox{${\it
PSL}(2,{\bf Z})$} containing~$H$. By the representation property
$N_H$ has then to be represented trivially, and nontrivial
representations, if existent, can only arise from $\hbox{${\it
PSL}(2,{\bf Z})$}/N_H$. Now notice that $A$ and $B$ each have
fixed points (for example $(1,0)$ and $(0,1)$, respectively), and
choose $H=\{A,B\}$. Since $\{A,B\}$ generates \hbox{${\it
PSL}(2,{\bf Z})$}, we have $N_H\cong \hbox{${\it PSL}(2,{\bf
Z})$}$.  We are thus left with only the trivial representation,
and we must take $\sigma_{(m,n)}=1$.

If we had allowed for orientation-reversing diffeomorphisms we
would have obtained $\hbox{${\it PGL}(2,{\bf
Z})$}\equiv\hbox{${\it GL}(2,{\bf Z})$}/\{1,-1\}$ instead of
\hbox{${\it PSL}(2,{\bf Z})$}. \hbox{${\it PGL}(2,{\bf Z})$} can
be generated by the three generators $T$, $S$, and~$R$, where
$R={\rm diag}(1,-1)$~\cite{coxeter}. But since the additional
generator, $R$, also has a fixed point it must be represented
trivially so that the case reduces to the one already discussed.

The feature that excludes all five non-trivial representations of
\hbox{${\it PSL}(2,{\bf Z})$} is the non-freeness of its action
on the set $\{{(m,n)}\}$, whose members indexed the classical
actions. Although we have here worked within the spatially flat
metric ansatz, this index set and the existence of the fixed
points have a purely topological origin: this index set labels the
different ways in which the manifold $T^2$ (on which the metric
appearing in the argument of the wave function is defined) can be
viewed as the boundary of the manifold ${\bar D}^2\times S^1$ (on
which the metrics contributing to the path integral were taken to
be defined~\cite{spatflat1}). The \hbox{${\it PSL}(2,{\bf Z})$}
action on our classical actions thus reflects the fact that the
diffeomorphisms of ${\bar D}^2\times S^1$ induce some, but not
all, of the disconnected diffeomorphisms of~$T^2$. We shall end
this section by demonstrating this.

Suppose we are given the two-manifold~$T^2$, and we want to view
it as a boundary of the three-manifold ${\bar D}^2\times S^1$. For
this we have to pick a simple, closed, oriented, non-contractible
curve, $\gamma_1$, on $T^2$ which is to be regarded as a boundary
curve $\partial\bar D^2\times q$ for some $q\in S^1$. We shall say
that its homotopy class, $[\gamma_1]$, has been canceled by
filling in a 2-disc. $\gamma_1$ can then be completed by another
such curve, $\gamma_2$, to form a generating basis,
$([\gamma_1],[\gamma_2])$ for the fundamental group
$\pi_1(T^2)={\bf Z}\times{\bf Z}$. With respect to a fiducial
basis, $([a],[b])$, of $\pi_1(T^2)$, these curves are uniquely
labeled by a matrix $M\in \hbox{${\it SL}(2,{\bf Z})$}$, i.e.
$([\gamma_1],[\gamma_2])=([a],[b])\cdot M$, so that the first
(second) column contains the winding numbers of $\gamma_1$
($\gamma_2$) with respect to $(a,b)$. Diffeomorphisms of ${\bar
D}^2\times S^1$ must act on $M$ by right multiplication with some
\hbox{${\it SL}(2,{\bf Z})$} matrix~$K$. It is shown in Appendix
\ref{app:isotopy} that $K$ must be generated by the matrices $-1$
and~$A$, and further that the set of $M$'s modulo right
multiplications by $K$ is in bijective correspondence to the set
of coprime integers modulo overall sign.

\section{3+1 dimensions}
\label{sec:3+1}

Among the ten closed spatial topologies that are compatible with
spatial flatness in 3+1 dimensions~\cite{wolf}, there are six for
which classical solutions  with the no-boundary data exist, and
among these six there are four for which the minisuperspace
spatial diffeomorphisms act nontrivially in the set of the
classical no-boundary actions~\cite{spatflat1}. We follow the
notation of Ref.~\cite{wolf}, and refer to these topologies as
${\cal G}_1$, ${\cal G}_2$, ${\cal B}_1$, and ${\cal B}_2$. We now
consider each of these topologies in turn.

\subsection{${\cal G}_1$ spatial surfaces ($S^1\times S^1\times
S^1$)}
\label{subsec:gone}

The simplest spatial topology is the three-torus
$S^1\times S^1\times S^1$, which in Refs.~\cite{spatflat1,wolf} is
referred to as~${\cal G}_1$. The identifications in the
coordinates $\{x^i\}=\{x,y,z\}$ are the periodic ones,
\FL
\begin{equation}
(x,y,z) \sim (x+2\pi,y,z) \sim (x,y+2\pi,z) \sim (x,y,z+2\pi)
\ \ ,
\label{goneidents}
\end{equation}
and ${h_{ij}}$ is a symmetric positive definite $3\times3$ matrix
with six independent components. We thus have $G=\hbox{${\it
GL}(3,{\bf Z})$}\cong \hbox{${\it SL}(3,{\bf Z})$}\times {\bf
Z}_2$, where the ${\bf Z}_2$ is generated by the orientation
reversing $-E={\rm diag}(-1,-1,-1)$ and \hbox{${\it SL}(3,{\bf
Z})$}\ by all the orientation preserving minisuperspace spatial
diffeomorphisms. In the action of \hbox{${\it GL}(3,{\bf Z})$}\ on
${h_{ij}}$ by~(\ref{minidiffeoh}) the only degeneracy is given by
the ${\bf Z}_2$-factor so that ${\tilde G}=\hbox{${\it SL}(3,{\bf
Z})$}$. We are thus automatically led to consider only
orientation-preserving diffeomorphisms. This also applies to the
other orientable case discussed under Section~\ref{subsec:gtwo}.

The actions of the classical solutions with the no-boundary data
are labelled by ordered triplets of coprime integers, modulo the
overall sign of the triplet. In analogy with
Section~\ref{sec:2+1}, we shall denote the equivalence classes of
such triplets modulo the overall sign by~${(m,n,p)}$. The
explicit formula is~\cite{spatflat1}
\widetext
\begin{equation}
I_{(m,n,p)} \left({h_{ij}}\right)
\>=\>
- {\pi^2 \over G}
\,
\sqrt{h\over
{m^2h_{xx}+
n^2h_{yy}+
p^2h_{zz}+
2mnh_{xy}+
2nph_{yz}+
2pmh_{zx}}
}
\label{Imnp}
\end{equation}
where $h=\det\left({h_{ij}}\right)$.
\narrowtext

On the set of the actions~(\ref{Imnp}), the \hbox{${\it
SL}(3,{\bf Z})$}\ transformations act in an obviously transitive
fashion analogous to that in~(\ref{Itransform}). The assumption of
linear independence then implies as in Section~\ref{sec:2+1} that
the  prefactors must take a form analogous to that in
(\ref{prefactorsplit}-\ref{sigmatransform}). The numbers
$\sigma_{(m,n,p)}$ are then fixed to be unity by the fact that
the only 1-dimensional representation of \hbox{${\it SL}(3,{\bf
Z})$}\ is the trivial representation. This follows from the
observation that the abelianization of \hbox{${\it SL}(3,{\bf
Z})$}\ is trivial, as may be explicitly checked from the
presentation 7.37 in Ref.~\cite{coxeter}.

\subsection{${\cal G}_2$ spatial surfaces}
\label{subsec:gtwo}

With ${\cal G}_2$ spatial surfaces the coordinates
$\{x^i\}=\{x,y,z\}$ have in addition to (\ref{goneidents}) one
further identification given by
\begin{equation}
(x,y,z) \sim (-x,-y,z+\pi)
\ \ ,
\label{gtwoidents}
\end{equation}
and the matrix ${h_{ij}}$ is constrained to satisfy
\begin{equation}
h_{xz}=h_{yz}=0
\ \ .
\label{gtwoconditions}
\end{equation}
Preservation of the identifications implies that the matrix $M$
must be of the block diagonal form
\begin{equation}
M \;=\;
\pmatrix{
p &r &0 \cr
q &s &0 \cr
0 &0 &\pm1 \cr}
\ \ .
\label{gtwoM}
\end{equation}
The spatial surface is orientable, and as in Section
\ref{subsec:gone} we can without loss of generality restrict the
attention to orientation-preserving minisuperspace spatial
diffeomorphisms and impose $\det(M)=1$. The orientable
minisuperspace spatial diffeomorphism group $G$ is therefore
(isomorphic to) \hbox{${\it GL}(2,{\bf Z})$}: the sign of the
$\pm1$ at the lower right corner in (\ref{gtwoM}) is equal to the
subdeterminant of the upper left $2\times2$ block. In the action
of $G$ on ${h_{ij}}$ by~(\ref{minidiffeoh}), the only degeneracy
is in the overall sign of the upper left $2\times2$ block (or
equivalently in the sign of the $\pm1$ at the lower right corner).
Hence ${\tilde G}=\hbox{${\it PGL}(2,{\bf Z})$}$. (Note that the
group \hbox{${\it PSL}(2,{\bf Z})$} quoted in
Ref.~\cite{spatflat1} is incorrect: it is only a subgroup
of~${\tilde G}$. Similarly, we shall see that the groups quoted in
Ref.~\cite{spatflat1} for the topologies ${\cal B}_1$ and ${\cal
B}_2$ are incorrect and in fact only subgroups of the
respective~${\tilde G}$'s.)

The actions of the classical solutions with the no-boundary data
are labelled by ordered pairs of coprime integers, modulo the
overall sign of the pair. We follow the notation of Section
\ref{sec:2+1} and denote the equivalence classes of such pairs
modulo the overall sign by~${(m,n)}$. The explicit formula for
the action is then obtained from (\ref{Imnp}) by setting $p=0$ and
multiplying the right hand side by~${1\over2}$.

The action of the \hbox{${\it PGL}(2,{\bf Z})$} transformations
on the set of the classical actions is similar to that
in~(\ref{Itransform}), where $M$ is now identified with the upper
left $2\times2$ block in~(\ref{gtwoM}). The obvious transitivity
of this action and the assumption of linear independence lead to
equations analogous to (\ref{prefactorsplit}-\ref{sigmatransform})
for the prefactors. To see what $\sigma_{(m,n)}$ are possible,
recall~\cite{coxeter} that \hbox{${\it GL}(2,{\bf Z})$} is
generated by the three matrices $\{A,B,C\}$, where $A$ and $B$ are
as in (\ref{sltwogenerators1}) and
\begin{equation}
C
=
\pmatrix{
1&0\cr
0&-1\cr}
\ \ .
\label{matrixC}
\end{equation}
These three matrices can thus be regarded as generators of
\hbox{${\it PGL}(2,{\bf Z})$} when understood modulo the overall
sign. The action of $C$ has fixed points at ${(m,n)}=(1,0)$ and
${(m,n)}=(0,1)$, and the action of $A$ and $B$ was seen to possess
fixed points in Section~\ref{sec:2+1}. This implies
$\chi(A)=\chi(B)=\chi(C)=1$, so the only possibility is the
trivial representation and $\sigma_{(m,n)}=1$.

Note that \hbox{${\it PGL}(2,{\bf Z})$} does possess three
nontrivial 1-dimensional representations~\cite{coxeter}. The
situation is analogous to that with \hbox{${\it PSL}(2,{\bf Z})$}
in Section~\ref{sec:2+1}: the nontrivial representations are only
ruled out by the fixed points in the \hbox{${\it PGL}(2,{\bf Z})$}
action on the semiclassical contributions.

\subsection{${\cal B}_1$ spatial surfaces}
\label{subsec:bone}

With ${\cal B}_1$ spatial surfaces the coordinates
$\{x^i\}=\{x,y,z\}$ have in addition to (\ref{goneidents}) one
further identification given by
\begin{equation}
(x,y,z) \sim (x+\pi,y,-z)
\ \ ,
\label{boneidents}
\end{equation}
and the matrix ${h_{ij}}$ is constrained by
satisfy~(\ref{gtwoconditions}). Preservation of the
identifications implies that the matrix $M$ must  be of the block
diagonal form~(\ref{gtwoM}), but with the additional condition
that the element $q$ must be even. As the spatial surface is not
orientable, both signs of $\det(M)$ must be allowed. The
minisuperspace spatial diffeomorphism group $G$ is therefore a
direct product of two factors, corresponding to the two blocks
in~(\ref{gtwoM}). The upper left $2\times2$ block corresponds to
the subgroup of \hbox{${\it GL}(2,{\bf Z})$} where the lower
non-diagonal element is even. The lower right $1\times1$ block
corresponds to the group~${\bf Z}_2$.

In the action of $G$ on ${h_{ij}}$ by~(\ref{minidiffeoh}) the
${\bf Z}_2$ factor is trivial, whereas the only degeneracy in the
other factor is in the overall sign of the  $2\times2$ matrix.
Hence $\tilde G$ is the subgroup of \hbox{${\it PGL}(2,{\bf Z})$}
where the lower nondiagonal element of (the equivalence classes
of) the  matrices is even.

The actions of the classical solutions with the no-boundary data
fall into two qualitatively different classes~\cite{spatflat1}.
In the first class there is just one action, obtained from
(\ref{Imnp}) by setting ${(m,n,p)}=(0,0,1)$ and  multiplying the
right hand side by~${1\over2}$. In the second class there are a
countable infinity of actions, labelled by ordered pairs of
coprime integers modulo the overall sign of the pair, with the
second integer odd. As above, we denote the equivalence classes of
such pairs modulo the overall sign by~${(m,n)}$, with now
$n$~odd. The explicit formula for the actions is then obtained
from (\ref{Imnp}) by setting $p=0$ and multiplying the right hand
side by~${1\over2}$.

The action with ${(m,n,p)}=(0,0,1)$ is clearly invariant
under~${\tilde G}$. If this action contributes to the wave
function with a nonzero prefactor, the wave function must
transform trivially.

Suppose then that the wave function only receives contributions
from the actions labelled by ${(m,n)}$ with $n$ odd. The $\tilde
G$ transformations act on this set of actions in a fashion
analogous to that in~(\ref{Itransform}), with $M$ being identified
with the upper left $2\times2$ block in~(\ref{gtwoM}). This action
is obviously transitive, which together with the assumption of
linear independence leads to relations similar
to~(\ref{prefactorsplit}-\ref{sigmatransform}).  We shall show
that there are exactly two possible representations.

Observe first that the subgroup of \hbox{${\it GL}(2,{\bf Z})$}
where the lower nondiagonal element is even is generated by the
three matrices $\{A,B^2,C\}$, where $A$, $B$ and $C$ are as above.
(This is seen for example by induction in the lower nondiagonal
element.) As above, we can regard these matrices as the generators
of $\tilde G$ when understood modulo the overall sign. The pair
${(m,n)}=(0,1)$ is a fixed point of both $B^2$ and $C$, which
implies $\chi(B^2)=\chi(C)=1$. We need to find the possible values
of~$\chi(A)$.

Let us find all one-dimensional representations of $\tilde G$ for
which $\chi(B^2)=\chi(C)=1$. The identity $ACAC=1$ implies
$\chi(A)=\epsilon$ with $\epsilon=\pm1$. Thus, in addition to the
trivial representation obtained with $\epsilon=1$, there can
exist at most one non-trivial representation. That such a
representation does exist can be shown by explicit construction:
the representations with $\epsilon=\pm1$ are given respectively by
\begin{equation}
\chi
\left(
\pmatrix{
p &r \cr
q &s \cr}
\right)
\;=\; \epsilon^r
\ \ .
\label{bonechi}
\end{equation}

What remains is to show that both signs of $\epsilon$ are
compatible with the transformation law (\ref{sigmatransform}) for
$\sigma_{(m,n)}$. To see this, observe that the parity of $m$ is
invariant under $B^2$ and $C$ but changes under~$A$. Compatibility
is therefore achieved by taking $\sigma_{(m,n)}=\epsilon^m$.

\subsection{${\cal B}_2$ spatial surfaces}
\label{subsec:btwo}

With ${\cal B}_2$ spatial surfaces the coordinates
$\{x^i\}=\{x,y,z\}$ have in addition to (\ref{goneidents}) one
further identification given by
\begin{equation}
(x,y,z) \sim (x+z+\pi,y+z,-z)
\ \ ,
\label{btwoidents}
\end{equation}
and the matrix ${h_{ij}}$ is constrained to satisfy
\begin{equation}
\begin{array}{rcl}
h_{xx}+h_{xy} &=& 2h_{xz} \\
h_{yy}+h_{xy} &=& 2h_{yz}
\ \ .
\end{array}
\label{btwoconditions}
\end{equation}
Preservation of the identifications implies that the matrix $M$
must be of the form
\begin{equation}
M \;=\;
\pmatrix{
p &r &(p+r\mp1)/2 \cr
q &s &(q+s\mp1)/2 \cr
0 &0 &\pm1 \cr}
\ \ ,
\label{btwoM}
\end{equation}
where now both $q$ and $r$ are even. As the spatial surface is not
orientable, both signs of $\det(M)$ must be allowed. From the
behaviour of matrices of this kind under matrix multiplication and
inversion it follows that the minisuperspace spatial
diffeomorphism group $G$ is again a direct product of two factors.
One factor is the subgroup of \hbox{${\it GL}(2,{\bf Z})$} where
both non-diagonal elements are even, coming from the upper left
$2\times2$ block in~(\ref{btwoM}). The other factor is~${\bf
Z}_2$, coming from the sign of $\pm1$ in~(\ref{btwoM}).

In the action of $G$ on ${h_{ij}}$ the ${\bf Z}_2$ factor is
trivial by virtue of the constraints~(\ref{btwoconditions}). In
the action of the other factor the only degeneracy is again in the
overall sign of the $2\times2$ matrix. Hence $\tilde G$ is the
subgroup of \hbox{${\it PGL}(2,{\bf Z})$} where both nondiagonal
elements of (the equivalence classes of) the  matrices are even.

The actions of the classical solutions with the no-boundary data
fall again into two qualitatively different
classes~\cite{spatflat1}. In the first class there is just one
action, obtained from (\ref{Imnp}) by setting
${(m,n,p)}=(1,1,-2)$ and multiplying the right hand side
by~${1\over2}$. In the second class there are a countable infinity
of actions, labelled by ordered pairs of coprime integers modulo
the overall sign of the pair, with now both integers odd. As
above, we denote the equivalence classes of such pairs modulo the
overall sign by~${(m,n)}$. The explicit formula for the action is
then obtained from (\ref{Imnp}) by setting $p=0$ and multiplying
the right hand side by~${1\over2}$.

The action with ${(m,n,p)}=(1,1,-2)$ is clearly invariant
under~${\tilde G}$. If this action contributes to the wave
function with a nonzero prefactor, the wave function must
transform trivially.

Suppose then that the wave function only receives contributions
from the actions labelled by~${(m,n)}$. The $\tilde G$
transformations act in a fashion analogous to that
in~(\ref{Itransform}), with $M$ being identified with the upper
left $2\times2$ block in~(\ref{btwoM}). The obvious transitivity
of this action and the assumption of linear independence lead to
relations similar to~(\ref{prefactorsplit}-\ref{sigmatransform}).
We shall show that there exist exactly two possible
representations.

Observe first that the subgroup of \hbox{${\it GL}(2,{\bf Z})$}
with both nondiagonal elements even is generated by the four
matrices $\{A^2,B^2,C,Z\}$~\cite{sanov}, or equivalently by the
four matrices $\{\alpha,\beta,C,Z\}$, where  $Z={\rm
diag\,}(-1,-1)$ and
\FL
\begin{equation}
\alpha = CA^2 =
\pmatrix{
1 &2 \cr
0 &-1 \cr },
\ \ \ \
\beta = ZCB^2 =
\pmatrix{
-1 &0 \cr
2 &1 \cr }
\ \ .
\label{alphabeta}
\end{equation}
As above, we can regard these matrices as the generators of
$\tilde G$ when understood modulo the overall sign. The pair
${(m,n)}=(1,-1)$ is a fixed point of both $\alpha$ and $\beta$,
which implies $\chi(\alpha)=\chi(\beta)=1$. $\chi(Z)=1$ follows
trivially from the definition of~$\tilde G$. We need to find the
possible values of $\chi(C)$.

Let us find all one-dimensional representations of $\tilde G$ for
which $\chi(\alpha)=\chi(\beta)=1$. The identity $C^2=1$ implies
$\chi(C)=\epsilon$ with $\epsilon=\pm1$. Representations with
both signs of $\epsilon$ exist by explicit construction: they are
given by
\begin{equation}
\chi
\left(
\pmatrix{
p &r \cr
q &s \cr}
\right)
\;=\; \epsilon^{{1\over2}(p + q + r + s)+1}
\ \ .
\label{btwochi}
\end{equation}
That (\ref{btwochi}) indeed is a representation of $\tilde G$ in
the nontrivial case $\epsilon=-1$ follows from observing that
$\chi(st)=\chi(ts)=\chi(s)\chi(t)$ for all
$t\in\{\alpha,\beta,C,Z\}$ and all $s$ in the subgroup of
\hbox{${\it GL}(2,{\bf Z})$} with nondiagonal elements even.

What remains is to show that both signs of $\epsilon$ are
compatible with the transformation law (\ref{sigmatransform}) for
$\sigma_{(m,n)}$. To see this, observe that the parity of the
integer $(m+n)/2$ is invariant under $\alpha$ and $\beta$
(and~$Z$) but changes under~$C$. Compatibility is therefore
achieved by taking $\sigma_{(m,n)}=\epsilon^{(m+n)/2}$.

\section{Summary and discussion}
\label{sec:summary}

We have discussed the existence of semiclassical no-boundary
theta-sectors in locally spatially homogeneous cosmological
models with flat spatial sections, both in 2+1 and 3+1 spacetime
dimensions, such that the metric ansatz was kept in its most
general form compatible with Hamiltonian minisuperspace dynamics.
It was shown that nontrivial no-boundary theta-sectors exist only
in two of the models, both of which are 3+1-dimensional and have
non-orientable spatial topology. In both of these models the
nontrivial no-boundary theta-sector is unique.

In 2+1 dimensions with toroidal spatial topology, we were able to
show that the non-existence of nontrivial no-boundary
theta-sectors had a topological origin which transcended our
having first arrived at this result within the minisuperspace
ansatz. We found this origin in the relationship between the
disconnected diffeomorphisms on the three-manifold ${\bar D}\times
S^1$ (on which the metrics contributing to the no-boundary wave
function were taken to live) and those on its boundary~$S^1\times
S^1$ (on which the argument of the wave function was taken to
live). There exist diffeomorphisms on $S^1\times S^1$ that cannot
be extended into ${\bar D}\times S^1$, and the necessary condition
given by Hartle and Witt \cite{hartle/witt} for the existence of
nontrivial no-boundary theta-sectors is thus satisfied; however,
the nontrivial theta-sectors are excluded by the intertwining of
the extendible and nonextendible diffeomorphisms. This gives an
example of a situation where the necessary condition of Hartle and
Witt is not a sufficient one.

It seems conceivable that our results in the 3+1-dimensional
models have a similar topological origin. Although we have not
pursued this question systematically, it should be noted that both
in 2+1 and 3+1 dimensions the only restrictions on our metric
ansatz came from the local homogeneity, the choice of the spatial
topology, and the consistency of Hamiltonian minisuperspace
dynamics. It would have been possible to introduce by hand further
symmetries for the minisuperspace metric and then to examine the
resulting models in their own right, but the results about the
existence of canonical theta-sectors and no-boundary theta-sectors
in such models would in general be different. An example is given
by the simple model in the introduction: this model is obtained
from that discussed in Section~\ref{sec:2+1} by restricting the
metric to be diagonal. It shows that further truncations can
induce further theta sectors not present in the larger theory.
This is also seen in the canonical quantization of the diagonal
3+1 torus~\cite{hajicek/bi}.

In all models where the actions of the classical no-boundary
solutions transformed nontrivially under the minisuperspace spatial
diffeomorphisms, the set of such solutions was countably infinite
and the minisuperspace spatial diffeomorphisms acted transitively
on this set. This enforced us to write the no-boundary wave
functions both in the trivial and nontrivial theta-sectors as
countably infinite sums of semiclassical contributions. At this
point it would certainly be desirable to exhibit a greater degree
of rigour in order to extend the meaning of the sum as well as its
relation to the Wheeler-DeWitt equation beyond the level of formal
arguments. Specification of the function space for the wave
function and suitable regularization conditions are amongst the
minimal requirements. A possible framework for this is the
observation that in our models the Wheeler-DeWitt equation is a
hyperbolic differential equation on an $n+1$-dimensional
``space-time," where the ``time" direction corresponds to the
volume of the three-metric, and the theta-structure lies entirely
in the $n$-dimensional ``spatial" sections. One could then follow
the standard Klein-Gordon treatment and require the wave function
to be square integrable over the ``spatial" sections. For the
restricted case of diagonal three-torus geometries, the dynamical
inequivalence of the canonical theta-sectors in this framework has
been discussed in Ref.~\cite{hajicek/bi}. For a discussion of the
two-torus case, see Ref.~\cite{carlip}

Since our aim was to discuss on specific models the general
features of the Hartle-Hawking proposal we content ourselves with
the general setting of formal sum-over-histories rather than, say,
attempting a Hilbert-space formulation. Eventually, however, one
has to enter the discussion of how physical predictions arise from
the wavefunction and then, as related issue, specify the function
space in which the wave function is to live. Only then could one
predict possibly different implications of theta-sectors.

\nonum
\section{Acknowledgments}

We would like to thank Abhay Ashtekar, John Friedman, Petr
Hajicek, Rafael Sorkin and Lee Smolin for discussions and
comments. This work was supported in part by the Natural Sciences
and Engineering Research Council of Canada, and by the National
Science Foundation under Grant No.\ PHY 86-12424.

\appendix{Linear independence}
\label{app:linear}

In this Appendix we show that any finite subset of the
no-boundary semiclassical contributions in the 2+1-dimensional
model is linearly independent.

Let
\FL
\begin{equation}
{\psi_{(m,n)}}({h_{ij}}) =
P_{(m,n)}\left({h_{ij}}\right)
\exp\left[-I_{(m,n)} \left({h_{ij}}\right) \right]
\,,
\label{psimn}
\end{equation}
where $I_{(m,n)}$ are as in~(\ref{Imn}), and $P_{(m,n)}$ are
nonvanishing functions that are slowly varying compared with the
exponential factors. Let $\cal S$ be a finite non-empty set of
pairs ${(m,n)}$ of coprime integers modulo the overall sign.
Suppose that
\begin{equation}
0 \;=
\sum_{{(m,n)}\in{\cal S}}
a_{(m,n)} {\psi_{(m,n)}}({h_{ij}})
\label{linear}
\end{equation}
for some complex numbers~$a_{(m,n)}$.
Let $(m_0,n_0) \in {\cal S}$, and choose
\FL
\begin{equation}
{h_{ij}} \;=\;
{\left[
\pmatrix{
r&-n_0\cr
s&m_0\cr}
\pmatrix{
\varepsilon&0\cr
0&\varepsilon^{-1}\cr}
\pmatrix{
r&s\cr
-n_0&m_0\cr}
\right]}_{ij}
\label{hchoice}
\end{equation}
where $\varepsilon>0$, and $r$ and $s$ are integers satisfying
$rm_0 + sn_0 = 1$. Then
\FL
\begin{equation}
I_{(m,n)} \left({h_{ij}}\right)
=
- {\pi \over 2G}
{1 \over
\sqrt{
\varepsilon {(rm + sn)}^2 +
\varepsilon^{-1} {(n_0m - m_0n)}^2
}}
\,.
\label{Imnhchoice}
\end{equation}
At the limit $\varepsilon\to0$,
$\left(\psi_{(m_0,n_0)}({h_{ij}})\right)
\!\!\big/\!\!
\left(\psi_{(m,n)}({h_{ij}})\right)$
diverges exponentially for all ${(m,n)}\neq(m_0,n_0)$. Therefore
(\ref{linear}) implies $a_{(m_0,n_0)}=0$. As $(m_0,n_0) \in {\cal
S}$ was arbitrary, the set
$\bigl\{{\psi_{(m,n)}}\bigm|{(m,n)}\in{\cal S}\bigr\}$ is linearly
independent.

\appendix{Isotopy classes of ${\bar D}^2\times S^1$ and $T^2$}
\label{app:isotopy}

In this appendix we discuss the isotopy classes of
diffeomorphisms of ${\cal M}=\bar D^2\times S^1$ versus those of
its boundary, $T^2$, and correspondingly show which boundary
diffeomorphisms are extendible to ${\cal M}$. We shall start with
a general setting and then specialize to the case under
consideration.

Let ${\cal M}$ be a manifold with connected boundary
$\partial{\cal M}$ and diffeomorphism groups $D({\cal M})$ and
$D(\partial {\cal M})$ respectively. Any $\phi\in D({\cal M})$
induces a $\partial\phi\in D(\partial {\cal M})$ by restriction:
$\partial\phi:=\phi\vert_{\partial {\cal M}}$. If
$\phi_0,\phi_1\in D({\cal M})$ are isotopic (are connected by a
one-parameter family of diffeomorphisms): $\phi_t\in D({\cal
M}),\,\forall t\in [0,1]$, then clearly $\partial\phi_0$ and
$\partial\phi_1$ are isotopically related by $\partial\phi_t$.
Equivalently, if two maps $\partial\phi_0,\partial\phi_1$ are not
isotopic in $D(\partial {\cal M})$, their extensions
$\phi_0,\phi_1$ are not isotopic in $D({\cal M})$ (note that this
would be false if we replaced isotopic by homotopic).  The
restriction map
\begin{equation}
\partial:\,D({\cal M})\rightarrow D(\partial {\cal M}),
\quad \phi\mapsto\partial\phi
\label{restrictionmap}
\end{equation}
thus projects to
\begin{equation}
\partial_*:\,H_{{\cal M}}\rightarrow
H_{\partial {\cal M}}, \quad [\phi]\mapsto
\partial_*[\phi]:=[\partial\phi]
\label{restprojection}
\end{equation}
where, if $D_0$ denotes the connected component of the
diffeomorphism group $D$, $H_{{\cal
M}}:=D({\cal M})/D_0({\cal M})=\pi_0(D({\cal M}))$ and
$H_{\partial {\cal M}}= D(\partial {\cal M})/D_0(\partial {\cal
M})=\pi_0(D(\partial {\cal M}))$ are the homeotopy groups of
${\cal M}$ and $\partial {\cal M}$. The image of $H_{{\cal M}}$
under $\partial_*$ in $H_{\partial {\cal M}}$, which we call $H'$,
is a subgroup of $H_{\partial {\cal M}}$ which, in general, is
properly contained. If $[\alpha]\in H_{\partial {\cal M}}-H'$
there is no diffeomorphism of ${\cal M}$ inducing $\alpha$; in
other words, $\alpha\in D(\partial {\cal M})$ can not be extended
to a diffeomorphism on ${\cal M}$. The set of non-extendible
diffeomorphisms of $\partial{\cal M}$ can conveniently be
identified with the coset space $H_{\partial {\cal M}}/H'$.

For the case in question: ${\cal M}=\bar D^2\times S^1$, $\partial
{\cal M}=T^2$, we know that $H_{T^2}\cong
Aut({\bf Z}\times{\bf Z})=\hbox{${\it GL}(2,{\bf Z})$}$. Let the
fundamental group of $T^2$ be generated by
$[\gamma_1],[\gamma_2]$, where $\gamma_1$ bounds a disc in ${\cal
M}$ (i.e. $\gamma_1$ is a meridian) which we denote by
$\gamma_1\sim 1$. Any diffeomorphism $\phi$ must act on these
generators by some \hbox{${\it GL}(2,{\bf Z})$}-matrix:
\begin{eqnarray}
([\gamma_1],[\gamma_2])\,\mathop{\longmapsto}^{\partial\phi_*}\,
&&([\gamma'_1], [\gamma'_2]):=([\gamma_1],[\gamma_2])
\cdot\pmatrix{a&b\cr c&d\cr}, \nonumber \\
ad-bc&&=\pm 1
\label{phiaction}
\end{eqnarray}
But clearly, any diffeomorphism $\partial\phi$ of $T^2$ that is
the restriction of a diffeomorphism $\phi$ of ${\cal M}$ must
send $\gamma_1\sim 1$ to $\gamma'_1\sim 1$. Hence $c=0$, and $H'$
is contained in the subgroup generated by the three matrices:
\FL
\begin{equation}
C=\pmatrix{1&0\cr 0&-1\cr},\quad -C=\pmatrix{-1&0\cr 0&1\cr},\quad
A=\pmatrix{1&1\cr 0&1\cr}
\label{genmatrices}
\end{equation}
On the other hand, it is easy to see that there are
diffeomorphisms $\phi_C,\phi_{-C},\phi_A$ of ${\cal M}$ whose
images under $\partial_*$ are just the generators $C,-C,A$
respectively. For example, if $(r,\theta,\varphi)$ coordinatize
${\cal M}$ in the standard fashion, where $\varphi=const.$ is a
meridian, take $\phi_C(r,\theta,\varphi)=(r,-\theta,\varphi)$,
$\phi_{-C}(r,\theta,\varphi)=(r,\theta,-\varphi)$ and
$\phi_A(r,\theta,\varphi)= (r,\theta+\varphi,\varphi)$. $H'$ is
therefore equal to the group generated by $C,-C,A$ which is
isomorphic to ${\bf Z}_2\times ({\bf Z}_2\times_s{\bf Z})$, where
the first ${\bf Z}_2$ is generated by $\phi_C\circ\phi_{-C}$, the
second ${\bf Z}_2$ by the orientation reversing $\phi_C$ and $A$
by the Dehn-twist, $\phi_A$, about a meridian. $\times_s$ denotes
the semidirect product.

Right multiplication of any matrix $M\in \hbox{${\it GL}(2,{\bf
Z})$}$ by $C$ or $-C$ changes the sign of the first or second
column. Right multiplication by $A^n$ adds $n$-times the first
column to the second. Given $M$, $\det M=\pm 1$ determines the
second column uniquely up to sign and adding multiples of the
first, that is, up to right multiplications by $-C$ and any
integer power of $A$. This shows that $\hbox{${\it GL}(2,{\bf
Z})$}/H'$ is in bijective correspondence to the set of possible
first columns up to overall sign, in other words, the set of
coprime integers modulo overall sign.

For orientation preserving diffeomorphisms we have \hbox{${\it
SL}(2,{\bf Z})$} replacing \hbox{${\it GL}(2,{\bf Z})$} and the
single $-E$ replacing $C,-C$, so that $H'\cong {\bf Z}_2\times{\bf
Z}$ generated by $\{-E,A\}$. But this time the second column is
uniquely determined by the first up to right-$A^n$ multiplications
so that again we have that $\hbox{${\it SL}(2,{\bf Z})$}/H'$ is in
bijective correspondence with the set of coprime integers modulo
overall sign.

\end{document}